\documentclass[11pt]{article}

\makeatletter
\def\@xfootnote[#1]{%
  \protected@xdef\@thefnmark{#1}%
  \@footnotemark\@footnotetext}
\makeatother
\usepackage{graphicx}
\usepackage[font={footnotesize}]{caption}
\begin{document}
\begin{centering}
{\Large{{\bf A Life in Statistical Mechanics}}}
\vskip 0.2 cm

{\large{\bf Part 1: From Chedar in Taceva, \\to Yeshiva University in New York}\footnote[*]{Article submitted to  \textit{European Physical Journal H}. \\The text presented here has been revised by the authors based on the original oral history interview conducted by Luisa Bonolis and recorded in Paris, France, 11--16 October 2014.}}

\vskip 0.5 cm
Joel L. Lebowitz\footnote{e-mail: lebowitz@math.rutgers.edu}

Departments of Mathematics and Physics, Rutgers, The State University, 110 Frelinghuysen Road, Piscataway, NJ 08854
\vskip 0.2 cm
Luisa Bonolis\footnote{e-mail: luisa.bonolis@roma1.infn.it, lbonolis@mpiwg-berlin.mpg.de}

Max Planck Institute for the History of Science, Boltzmannstrasse 22, 14195 Berlin, Germany.
\end{centering}
\vskip 0.5 cm
\abstract{
This is the first part of an oral history interview on the lifelong involvement of Joel Lebowitz in the  development of statistical mechanics. Here the covered topics include the formative years, which overlapped the tragic period of Nazi power and World War II in Europe, the emigration to the United States in 1946 and the schooling there. It also includes the beginnings and early scientific works with Peter Bergmann, Oliver Penrose and many others. The second part will appear in a forthcoming issue of EPJ-H.
} %end of abstract

\section{From war ravaged Europe to New York}

L. B. \hspace{0.2 cm}
Let's start from the very beginning\dots Where were you born? 
\vskip 0.3 cm

J. L. \hspace{0.2 cm}
I was born in Taceva, a small town in the Carpathian mountains, in an
area which was at that time part of Czechoslovakia, on the border of Romania and
about a hundred kilometers from Poland. That was in 1930, and the town then had a
population of about ten thousand people, a small town, but fairly advanced. We had
electricity, which was not present in any of the neighboring villages. The population
was mostly Hungarian. I would say probably 50\% were Hungarian, 20\% were Jewish
and 30\% were Ruthenian. They spoke Ruthenian, a Slavic language, although some
say it is just a dialect of Ukrainian, while others claim it is a separate language. But
anyway, when I was growing up, my mother tongue was Yiddish.

Taceva was part of Czechoslovakia but there were very few Czechs there. In the
whole town there were just about ten or twenty Czechs: the police chief and perhaps
some other officials of the town. Czechoslovakia was created after the First World War, when they divided up the Austro-Hungarian empire. It consisted then of Bohemia,
Moravia, Slovakia and my part, which was called Carpathia. In my village, there
was an intermingled population, which, as far as commercial relations existed and
as far as where people lived, was totally mixed. But where social interactions were
concerned, the Jewish population was totally separate and so were the Hungarians
and the Ruthenians. They each spoke different languages though, of course, being
a minority, the Jewish population certainly had to learn to speak Hungarian also.
Before the First World War, our area had been in the Hungarian part of the Austro-
Hungarian empire. So Hungarian was then the official language.
\vskip 0.3 cm

L. B. \hspace{0.2 cm}
And your parents? Where did they come from?
\vskip 0.3 cm

J. L. \hspace{0.2 cm}
My mother was born in the same village, and her parents --- my grandparents
--- still lived there, as well as two of her sisters, while a brother had moved about
fifty kilometers away. My father came from a slightly bigger town, named Chust,
which was about thirty kilometers away. His mother and several of his siblings still
lived in that town. When my father was about twelve years old, his father emigrated
to America. There are stories in the family that my grandmother did not consider
him a good role model for their eight children (he apparently liked to gamble!). So I
never knew him at all, only my grandmother. Going to visit my grandmother was a
big event for me, as a little boy, although in reality it was only about thirty kilometers
away. But whenever you went from Taceva to Chust, you would be given lots of food
to take along to eat on the train, since this was considered a big journey even though
it took only about an hour.

At the age of three, like almost every other Jewish boy, I started going to Cheder.
Cheder usually lasted from the ages of three to about thirteen or fourteen. Afterwards
one went to a Yeshiva for further talmudic study. Cheder was a Jewish school and
it was for young boys only. There was already some kind of gender differentiation.
Up to the age of three, little boys and little girls wore their hair in the same style,
just as it grew naturally. But at the age of three, the boys' long hair was cut short,
sparing only their payes, or sidelocks. I think I started going to Cheder soon after
my haircut, as was typical for Jewish boys. Then at the age of six one also had to
attend a regular state primary school (though in my case it was postponed to age
seven). This school, the regular state school, was secular. But we did not give up our
religious training, we just changed the schedule. So, at age six or seven, we began to
go to Cheder very early in the morning at six-thirty or something like that. We came
home at maybe eight or eight-thirty, had breakfast, and then went to the secular
school and took classes until about two o'clock in the afternoon, and then went back
again to the Jewish school, after having stopped at home for lunch.

In my first year at the state school, classes were taught in Czech, because Taceva
was part of Czechoslovakia at that time. Everything was in Czech, even though none of
the students could speak Czech because they had Yiddish or Hungarian or Ruthenian
backgrounds.

And then, in my second year at school --- it was in the fall of 1938 --- we
started out as usual in the Czech school, but then Germany took over Bohemia and
Moravia and installed a puppet government in Slovakia, and in our area a kind of a
Nazi ideological group of Ruthenians took over. So the school switched from being
taught in Czech to being taught in Ruthenian. Then, after three months, it switched
again when the Hungarians took over. I continued in the Hungarian secular school
with the same schedule as before, until I finished the compulsory 6th grade. In 1943,
I guess. Then, in order to spend more time on Jewish studies, I took an exam as a
substitute for continuing in school for another two years.
\vskip 0.3 cm

L. B. \hspace{0.2 cm}
What kind of subjects did you study at school?
\vskip 0.3 cm

J. L. \hspace{0.2 cm}
In the Hungarian school we studied a lot of Hungarian geography. At
an exam you had to stand with your back to a map of Hungary and be able to
point out the different cities. We also had Hungarian grammar and literature and
some mathematics. Perhaps some physics, too. I remember a teacher showing us a
ring and a ball. The ball would go through the ring when it was cold, but after he
heated it up with a match it would not. This demonstrated thermal expansion. I
still remember it. We did not do any experimental science. I guess I was quite good
at mathematics. The teacher would ask a question. I would yell out the answer. All
these secular subjects were considered in my group as a kind of necessary evil. We
were then mainly focused on Jewish studies, but still you had to attend the secular
state school. I think that there was some appreciation even in our Jewish group
that for everyday living you had to know some non-Jewish subjects. We also had
some reading, of Hungarian literature, as well as some fairytales, like those of Hans
Christian Andersen translated into Hungarian. I do not think there was anything like
that available to us in Yiddish. I think everything was in Hungarian. So, from the
age of eight until the age of thirteen, I went to this Hungarian school.

Thirteen, in the Jewish calendar, is an important birthday. It is then that one
graduates from being a boy and becomes a man. There are certain religious ceremonies
and requirements. So then I stopped going to the Hungarian secular school. Instead,
I went to do more Jewish studies at a Yeshiva, which was in a village called Hadasz.
I studied at the Yeshiva, lived in a rented room with other students and ate there
for one semester and then --- I guess it was the Passover holiday of 1944, as I was
back at home in Taceva --- the Hungarian authorities, now under more direct German
control, proclaimed that all the Jews had to move to one side of the town. Then, after
a few weeks, we had to move again, this time to a particular neighborhood which was
then surrounded by wire fences and became a ghetto. And then the nightmare really
started. That was the beginning of the deportations of the Jews in our area to the
concentration camps. So there I was, in the ghetto, from April to May, when I turned
fourteen.

The trip to the concentration camp was done by train in crowded cattle cars. The
trip was horrible: almost no food or water, no toilet facilities. After about a week of
travel in this car, we arrived at Auschwitz on May 26, 1944. There the doors were
opened and we were pushed out of the cars amid much shouting and shoving by 
commandos, special groups made up of selected camp inmates. Outside the cars the
Gestapo were pushing women, children, and old people into one line and healthy
looking men to another line. I followed my father into that line. There was a Gestapo
officer, I believe it was the nefarious Dr. Mengele, who inspected everyone, sending
those that looked weak to the other side. When I came up to him he looked at me and
asked how old I was. Some instinct told me to lie and say I was fifteen. So I passed.
Had I been sent to the other side I would most probably have been taken directly,
together with my mother, sister, grandmother and aunts, to certain death in the gas
chamber.

I was in the concentration camps from age fourteen to fifteen. By the time Germany
was defeated in May, 1945, most of my family was dead. All my immediate
family was gone: my mother, my father, my sister, and my grandmothers. They were
all murdered in the camps. I will not describe here my experiences in the concentration
camps. They were similar in nature to those described by Elie Wiesel, Primo
Levi and Imre Kertesz.

After my liberation by British troops from the concentration camp at Bergen-
Belsen in May, 1945, I and other former inmates wandered around the area, begging
and extorting food from the local population. We were quartered in a kind of a
rehabilitation camp for a few weeks, and then loaded on buses and repatriated to
(then Russian-occupied) Czechoslovakia. The buses dropped us  at some train
station. From there we traveled by train.

After many stops and some unpleasant encounters with the occupying Soviet
troops, we arrived in Budapest, where we were sheltered and given some money
by a Jewish-American aid society, nick-named the ``Joint'' (American Jewish Joint
Distribution Committee). There I met a surviving uncle, and found out about who
else in my extended family might have survived. I stayed in Budapest for a week or
so. From there I went back by train to my home in Taceva, hoping to  find some family
silver and other things of value that we had hidden there when we were deported. My
quest was unsuccessful, however; everything was gone.

There was almost nobody I knew left in Taceva. But I had an aunt, my father's
sister, who had formerly lived in Chust, the place where my father was born. She and
two of her children had survived and she had moved with them to Szatmar, a city in
Romania that was close to Taceva.

When I grew up, Taceva was in Czechoslovakia, but it was right at the Romanian
border. As children, we would go down a winding path to the edge of the river, and
look across the water to Romania. We would also go shopping there for halva. Then,
in about 1941, Hungary took over that part of Romania, too. After WWII, in 1945,
it reverted to being Romania. But the place where I was born, Taceva, became part
of Ukraine, (which was then part of the Soviet Union). I stealthily crossed the border
and moved in with my aunt for a while. This was all in the summer of '45. In the fall,
when my aunt and her children left Szatmar, I stayed in Romania and began going
again to a Yeshiva, where I had no secular studies at all.

In the spring of '46 --- I guess it must have been around March or April --- I decided
to leave the Yeshiva and rejoin my aunt who was now living in Czechoslovakia (now
Czechia) near Prague. To do that I had to cross several borders, so there were all
kinds of problems and difficulties. But eventually I managed to get to Prague, where
somebody from my hometown was organizing (illegal) transportation for uprooted
and displaced people like me to go to Belgium. We were put on a bus and equipped
with some false travel documents. The idea was to go to Belgium and learn to work in
the diamond-polishing trade, and then to go on to Palestine. (The state of Israel had
not yet been created). Many people who had survived the death camps were going
there.

Belgium's economy was good then, because they had extra income from their
African colony of the Congo. I do not know who started it, but some Jewish organizations
had set up an arrangement so that we could get job training in Belgium.

\begin{figure}[ht]
\captionsetup{width=0.8\textwidth}
\centering
\includegraphics[scale=0.35]{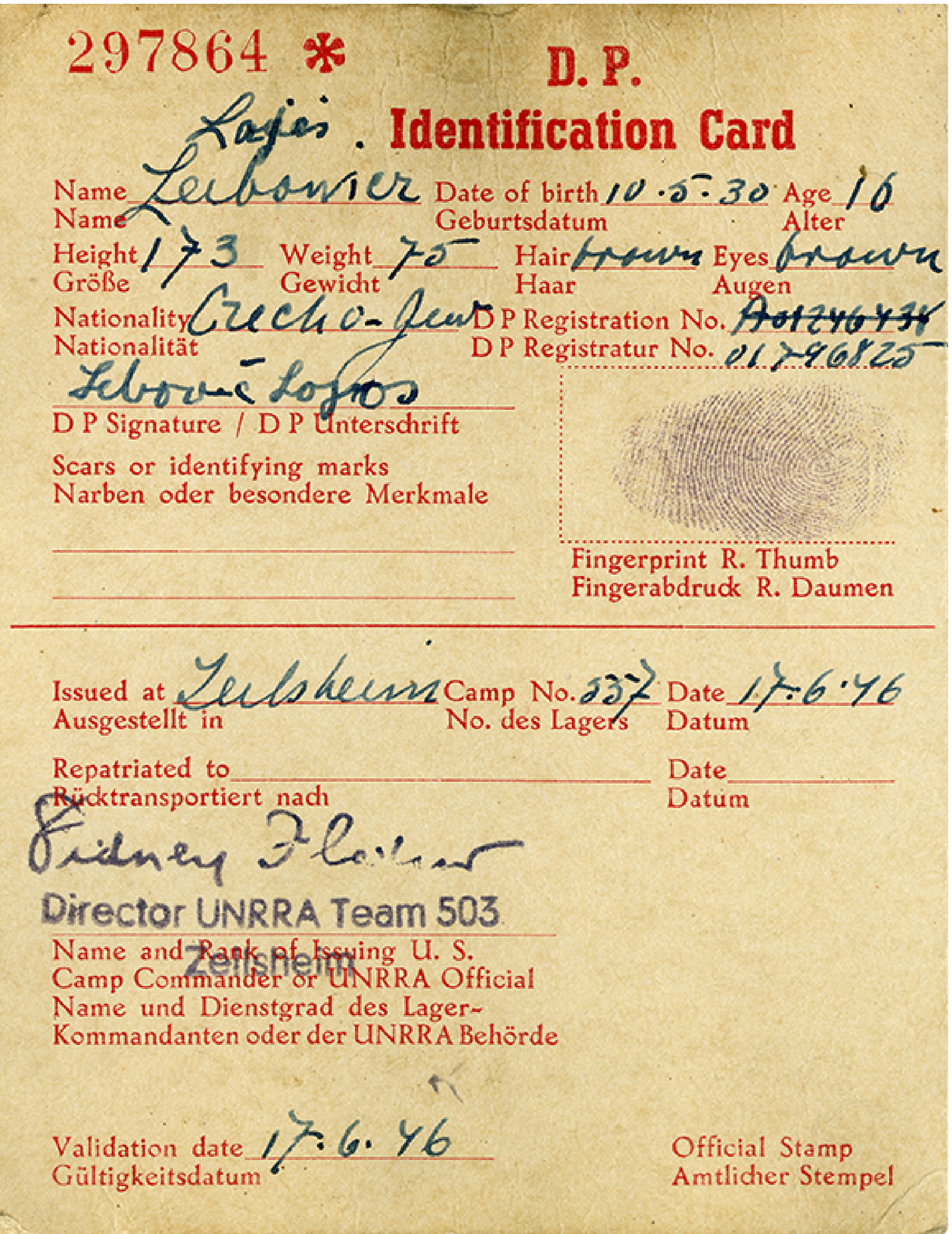}
%\resizebox{0.75\columnwidth}{!}{%
%  \includegraphics{fig1.eps} }
\caption{Lojos (Joel) Lebowitz's Displaced Person Identification Card issued at Zeilsheim Displaced Persons Camp on 17 June 1946. During this period there were about 3,500 people,  mostly survivors of concentration camps, housed in this camp located  in a suburb of Frankfurt am Main, in the American-occupied zone. They lived in small houses in an urban settlement once built by IG Farben for the company's workers. In 1946 it was visited by Eleanor Roosevelt and David Ben-Gurion. Most of the people living at Zeilsheim left Europe for either the United States or Palestine between 1945 and 1949.}
\label{fig:Dirac}       % Give a unique label
\end{figure}

However, our group never got there. Stopped at the border between Luxembourg and
Belgium, we were sent back to the adjacent French-occupied part of Germany. We
stayed there for a month in a now empty prisoner-of-war camp and eventually were
taken to the American-occupied part of Germany, where there were some camps for
displaced persons, run by a US government organization (mostly the Army) under
the auspices of UNRRA (United Nations Relief and Rehabilitation Administration).
I went to one of those camps located at Zeilsheim, very near Frankfurt. Then, from
there, at the end of August 1946, I emigrated to the United States.

\vskip 0.3 cm
L. B. \hspace{0.2 cm}
How did it happen?
\vskip 0.3 cm

J. L. \hspace{0.2 cm}
I was involved in a rescue project promoted by Eleanor Roosevelt. This
organization was active in helping orphaned children in the displaced persons camps
in Germany. These displaced persons camps after the war were temporary havens,
stopping-off places for people like me who were hoping to emigrate either to the United
States or to Palestine. But, actually, emigrating to Palestine was a risky business. The British did not permit immigration into Palestine at that time, so it had to be done
illegally. So applying to go to the United States was a better option for people like
me. I got a visa through one of Mrs. Roosevelt's projects, the U.S. Committee for the
Care of European Children (CARE), and I left Europe just at the end of August of
1946.

\section{Settling in the United States}

L. B. \hspace{0.2 cm}
What were your feelings when you left Europe? And your impressions upon
your arrival in the United States?
\vskip 0.3 cm

J. L. \hspace{0.2 cm}
I guess I was both apprehensive and glad to leave Europe where there
did not seem to be any future for me. With many other children in similar plights, I
was taken by CARE across the Atlantic to New York City. There I had an uncle, my
father's brother, who had come to the United States long ago, but I had not really
been in contact with him directly --- only through my aunt, his sister. I was still in
the charge of the people from the CARE, who first took the children, myself included,
from the boat to a shelter in the Bronx. From there I contacted my uncle, who came
to see me with his children, my American cousins.

A family I had met on the boat, the S. S. Perch, put me in contact with a Yeshiva
in Brooklyn, Yeshiva Torah Vodath, where I could have free room and board and
could complete my Jewish studies. To my surprise, this Yeshiva also had a regular
high school. It had a double program: a program in Jewish Talmudic studies in the
morning, and also a program of regular secular high-school studies in the afternoon.
In the spring semester of 1947, after some intensive classes  in English and Mathematics (arranged by the Yeshiva during the fall of '46), I joined in the regular secular
high-school program of the Yeshiva. There I studied Mathematics, Physics, History,
English literature, and Hebrew as a language. But also, to catch up, I went to evening
school, studying typing and German. I graduated from high school in July 1948. This
usually takes four years, but it was compressed into a year and a half --- that's why
I had to go to evening school to catch up with that! By 1948 I was not interested
anymore in studying in the religious part of the Yeshiva program. I was thinking, in
fact, of leaving the Yeshiva and getting a job and being more independent.
\vskip 0.3 cm

L. B. \hspace{0.2 cm}
You were still living at the Yeshiva?
\vskip 0.3 cm

J. L. \hspace{0.2 cm}
Yes, I lived in the dormitories there from the fall of 1946 until I graduated
from high school in 1948. And then I moved out from the dormitories and started to
work in a factory, a small factory that assembled cigarette lighters. My job there was
inserting the little wicks into the cigarette lighters. I did that for about two or three
weeks, but I was not very good at it and so I got fired after a few weeks.
\vskip 0.3 cm

L. B. \hspace{0.2 cm}
But this gave you the opportunity to earn some money so that you could
live somewhere else, not at the Yeshiva?
\vskip 0.3 cm

J. L. \hspace{0.2 cm}
Yes, exactly. But the head of the Yeshiva wanted me to return, hoping to
convert me back to religion. So I did go back and stayed there for another year, from
1948 to 1949. But I didn't live in the dormitory anymore. I had rented a room nearby
and lived there. In fact, I was also going to Brooklyn College in the evenings. This was
part of the City College of New York, now CUNY. I took the usual courses in History,
English, Mathematics and Physics. But from 1948 to 1949 I was doing that just in
the evening, and was still officially enrolled at the Yeshiva where I studied Talmud
during the daytime. I was again being supported by the Yeshiva, but it was really a necessity, as I was not working anymore. I did some tutoring or something like that
to make a little extra money. In fact, maybe in 1948 the children's aid organization
still gave me a small subsidy, not a large amount. I can't remember precisely. Then,
in 1949 I decided to leave the Yeshiva for good and go to Brooklyn College full time.
Of course, that meant I needed a job, and I started being a waiter in the evening, at
first in a deli-restaurant in the Polish section of Brooklyn. Then I got a better job ---
it must have been in the summer of 1950 { through a cousin of mine who used to eat
in a kosher restaurant in the Lower East Side of Manhattan. I worked at this Tel Aviv
Restaurant as a waiter five days a week. I would work there on Sunday afternoon from
3 to 8 o'clock, and then on Monday, Tuesday, Wednesday and Thursday from 5 to 8.
The people who owned the restaurant did the cooking and serving and everything.
They were also refugees who had come to the U.S. from Europe after the war. They
felt that taking tips was not dignified. There was a sign that said ``No Tips'' and so, at
first, I did not get any. But slowly I managed to cover up the ``No'' part and just left
``Tips''. I got a rather minimal salary but I also had my dinner at the restaurant, and
with my tips, too, I survived until I  finished college in 1952. By that time I actually
had savings of maybe  five hundred dollars or something like that, and I was even able
to buy a car for fifty dollars --- a used car, but not bad.

\section{Studying with Melba Phillips and Peter Bergmann}

L. B. \hspace{0.2 cm}
What do you remember about your studies at Brooklyn College?
\vskip 0.3 cm

J. L. \hspace{0.2 cm}
I came in contact there with some very good teachers, in particular the
physics teacher, Melba Phillips. She wrote a famous book, together with Wolfgang K. H. Panofski,
on electricity and magnetism \cite{PanofskiPhillips1955}. She had been a student of
Robert Oppenheimer --- there is a famous Oppenheimer-Phillips effect. During the
depression in the 1930s Brooklyn College had been able to get some very, very good
people, really excellent teachers. There were three or four outstanding mathematics
and physics teachers, particularly Melba Phillips. So, while on entering college I had
felt more like just doing engineering --- a practical thing --- I was soon converted
to theoretical physics. On graduating from College, I was accepted by various good
universities for graduate work.

I decided to go to Syracuse University because Peter G. Bergmann was there.
Bergmann worked in general relativity --- he had been a post-doc of Einstein after
he came to the United States in the 1930s. At some point he became interested in
statistical mechanics. He wrote several books on theoretical physics, one of them on
relativity, one on electromagnetism, but also one on statistical mechanics. He was a
friend of Melba Phillips, who recommended him very strongly, and so I decided to go
to Syracuse specifically to work with him in statistical mechanics.
\vskip 0.3 cm

L. B. \hspace{0.2 cm}
How did you discover Statistical Physics?
\vskip 0.3 cm

J. L. \hspace{0.2 cm}
Melba Phillips gave a course in statistical mechanics in my last year in
Brooklyn College and she used Peter Bergmann's book as a textbook. So I was studying
his book on statistical mechanics in the summer of 1952 before going to graduate
school. I had finished my undergraduate studies and graduated from Brooklyn College,
and was still working as a waiter in the Tel Aviv Restaurant in Manhattan
while preparing for graduate school. And I had gotten my  first car. I think I was also reading some other books on statistical mechanics, like Khinchin's book on the foundations of statistical mechanics \cite{Khinchin1949}. It certainly was a fascinating subject: how to describe macroscopic phenomena in terms of their microscopic constituents.
And I was also attracted by the probabilistic aspects of that. I had taken a
course in college on probability theory and found that statistical mechanics combines
both probabilistic and physical parts in a very nice way.

Once I got involved, I was quite interested in the subject both from a physical and
a mathematical point of view and so I decided to study with Bergmann. Bergmann
was still working mostly in general relativity and had been one of the few collaborators
of Einstein. He was a leading  figure in the field and had many visitors.

At that time Peter had many graduate students who were doing relativity. These
included Joshua Goldberg, Jim Anderson and Ralph Schiller, all of whom got their
degrees just before I got there. The only relativity student I was good friends with
during my period there is Ted Newman, of the Kerr-Newman metric. Besides me,
Bergmann had two other students, Charles Willis and Alice Thomson, who were
doing statistical mechanics.

I should mention as a background that Peter Bergmann was living in New York
City while he was teaching at Syracuse. This was for family reasons; his wife had a job
as a crystallographer at the Brooklyn Polytechnic Institute in New York (now part
of NYU) and Peter also had some part-time job there. He would take the overnight
train every week on Tuesday evening and be in Syracuse on Wednesday, Thursday
and Friday and then go back to New York. So I had my first meeting with him in his apartment in New York City. I liked him immediately | he was very personable | and that made me decide to go to Syracuse.

In Syracuse, I was able to get a Research Assistantship (paid from Peter's Air
Force grant), so I did not have to teach and I was able to survive as a graduate
student with this arrangement.

\section{Doing research with Melvin Lax and Eugene Gross}

I did my master's thesis with Melvin Lax, who was then a young assistant professor
at Syracuse. He gave me a problem having to do with the frequency distribution
$g(\omega)$ of harmonic crystals, using some new work by Leon Van Hove \cite{VanHove1953}.
A harmonic crystal is described by a quadratic Hamiltonian whose diagonalization
yields its fundamental frequencies $\omega({\bf k})$ where ${\bf k}$ is a wave vector specifying a normal mode. These {\bf k}'s have a uniform density inside a Brillouin zone. (In the simplest one-dimensional case this is just the interval, or circle [$-\pi,\pi$]. The problem then is to
get the fraction of frequencies, $g(\omega)$, in an interval $d\omega$.

Van Hove proved that this function will have singularities because the frequency
as a function of the normal modes $\omega({\bf k})$ is periodic, so the denominator of $g(\omega)$ which contains derivatives of $\omega({\bf k})$ has to have zeros, because it's a periodic function. (A smooth periodic function which is not a constant must have a maximum and a minimum
where the derivatives vanish.) Combining this information with knowledge of
the moments of $g(\omega)$, which it was also possible to compute, led to an approximation
which had the first correct 3 moments and at the same time had the right singularities.
So that was my very first published paper \cite{LaxLebowitz1954} and it was also my master's thesis.

There was also another young professor at Syracuse, Eugene P. Gross. You may
have heard of the Gross-Pitaevskii equation | that is an equation in super
fluidity. With Gross I did a problem on the time evolution of the density matrix for a dipole
immersed in a noble gas of neutral particles. We wanted to describe a situation where,
in analogy with classical mechanics, collisions change the velocities but do not change
the positions. To do this we made a certain assumption about the structure of the
time evolution of the density matrix of the system, which seemed reasonable. We
did not know at that time that quantum Markov evolutions have to have a certain
structure (the Lindblad condition), which ours did not have. So that means that our
results can only be approximate \cite{GrossLebowitz1956}.

\section{PhD Thesis}

L. B. \hspace{0.2 cm}
Your work with Gross was published in 1956. By that time you were
completing your PhD thesis under the supervision of Peter Bergmann: Statistical
mechanics of non-equilibrium processes. A very challenging and foundational work.
It appears that Bergmann, in parallel with his well established research activity on
relativity, had been interested in statistical physics at least since 1951, when he published
the article ``Generalized Statistical Mechanics'' \cite{Bergmann1951}. A hint of his
motivation for this work is provided in the first lines of the abstract: ``In this paper,
a start is made toward the development of a statistical mechanics that will be suited
to the treatment of dynamic changes in thermodynamic systems and which, at the
same time, will be in the appropriate form for a relativistically invariant theory.''
And then, in 1953, he published with Alice C. Thomson ``Generalized Statistical Mechanics and the Onsager Relations''  \cite{BergmannThomson1953}, where he is exploring
an approach using non-equilibrium ensembles. So with your thesis you should have
tackled the problem of non-equilibrium from a quite general point of view\dots
\vskip 0.3 cm

J. L. \hspace{0.2 cm}
Actually, there was relatively little work on non-equilibrium statistical
mechanics that I was aware of at that time.

I was particularly interested in obtaining mathematically exact solutions to precisely
posed problems, even if highly idealized from a physical point of view. I was
totally unaware then of the work in the Soviet Union by Bogoliubov, Kolmogorov
and their students. I was more aware of the rigorous results in equilibrium statistical
mechanics such as the Onsager solution of the Ising model, the Lee-Yang theorems
on phase transitions, the Berlin-Kac solution of the spherical model and van Hove's
results mentioned earlier. There was however no one at Syracuse pursuing this line
of research. 

I did become aware in the course of doing my thesis, of the works of Prigogine,
van Kampen, Onsager and van Hove on non-equilibrium.

Going back to my thesis, I do not remember seeing any work on open systems
in contact with several reservoirs. It seemed like a very natural question and still
seems very natural: systems which are not in equilibrium but are in a stationary
state maintained by contact with heat baths at different temperatures. I found this
an interesting problem to work on. It involved some serious mathematical questions
so I took some advanced courses in mathematics when I was in graduate school. I
remember in particular a course I took with a famous probabilist who was at Syracuse
University at that time, named Kay Lai Chung (who later went to Stanford). I was
also reading books on probability, particularly a fairly new book by Joseph L. Doob \cite{Doob1953}, a well known probabilist. There I found some results which seemed directly applicable to
the question I was studying, namely, finding sufficient conditions for a system in contact with reservoirs to approach a stationary state.

That stationary state would be an equilibrium state if there was only one reservoir
or if the temperatures of the different reservoirs were the same. In that case you
would know actually what that stationary state is. But in the case where the temperatures
of the reservoirs are different, the stationary state would be a non-equilibrium
one. I guess now it is usual to refer to such a state as a non-equilibrium steady state
(NESS). Doob was, in fact, quoting some results by a French-German mathematician,
Wolfgang D\"oblin, which seemed to address that question directly \cite{Doblin1937}. There is
something called the D\"oblin condition, saying that if you have a stochastic evolution such that the probability of the state of the system, $X$, going from any initial state in a given time $t$ into
a region of very small measure $\delta$, is less than $1-\delta$, then the system will approach a stationary state. If you had a deterministic evolution, then starting from a given initial state you will go after a time $t$ just to some very definite new state. But in a stochastic evolution you typically spread out so if the probability of going into a very small set becomes less than 1, then indeed you have an approach to a stationary state, in fact an exponential in time approach without having to know
what the stationary state is. I have very recently again used the results in my thesis \cite{CarlenEtAl2016}.

There was at that time some work, much more physical and approximate, on
how to model the reservoir but nothing mathematically rigorous. Physically what
happens when a system is coupled to a reservoir is that the information in the part of
the reservoir interacting with the system travels away very fast because the reservoir
has a very high heat conductivity. What we did --- I am not sure if we were the first
ones or not --- was to idealize the reservoir by saying that it acts on the system in a
stationary stochastic way. That is, we idealized the reservoir as consisting of a gas of
particles which come in from infinity, interact with the system, and then go away to
infinity and do not come back again. So you have a steady, time independent stream
of particles coming in, colliding with the system, and giving a transition probability
from one state of the system to another. A state of a classical system is defined as
a point in the phase space of the system. This means specifying the position and
velocity of all the particles in the system.

If the system is isolated it would have a deterministic evolution according to the
Hamiltonian equations of motion. But because of the impulsive interaction with the
reservoir, the phase points can jump, i. e. make a transition from one phase point to
another phase point, with a certain probability and this will happen independently
for the different reservoirs. So, as already mentioned, if you only had one reservoir at
a given temperature then eventually there would be a stationary state which is an
equilibrium Gibbs probability distribution at the temperature of that reservoir. This
state would satisfy a detailed balance condition. But, if you have several reservoirs
at different temperatures you get a competition. Each reservoir tries to drive the system
to a Gibbs distribution at its own temperature. The final stationary state will, in
general, be unknown but nevertheless one could show that this stationary state had
certain properties. In particular it would satisfy the Onsager relations, that is, the
dependence of the flux from reservoir $i$ into the system on the temperature of reservoir
$j$ is the same, in the linear regime, as the dependence of the flux from reservoir $j$ on
the temperature of reservoir $i$. You have a certain kind of symmetry that goes under
the name of Onsager relations. We were able to prove that these stationary states
satisfy these relations.

I finished my thesis in 1956 and wrote two joint papers with Peter Bergmann, which I still often refer to \cite{BergmannLebowitz1955} \cite{LebowitzBergmann1957}.

\section{Post-Doctoral Work}
In the summer of 1956, after I got my PhD but before starting my post-doc with
Onsager at Sterling Laboratory, Yale University (see below), I got a summer job
with an IBM theoretical group in New York. There I worked on transport properties
of semi-conductors in a group headed by Peter Price. I got the job partly because,
during the academic year 1955-1956 I stayed in New York a lot, for family reasons,
and hung out at Columbia University, where I met Peter Linhart, who was working
for Peter Price in this lab. I found the work there interesting, but it did not lead
to any publications. While I was working at the IBM lab in New York, I got a call
from someone at the Air Force Office of Scientific Research (AFOSR) saying that
Ilya Prigogine was organizing an international conference in statistical mechanics in
Brussels. It was one of those early IUPAP conferences, and I was invited. The AFOSR
was supporting that conference, so I could fly to Europe with what was called at
that time MATS (Military Air Transport Service). That was my first trip back to Europe, after having come to the United States in 1946. There were several other US
scientists going the same way to that conference. We flew to Paris making stops in Newfoundland and Ireland.

\vskip 0.3 cm

L. B. \hspace{0.2 cm}
How did you feel, coming back to Europe? But at the same time it was a
new experience for you!
\vskip 0.3 cm

J. L. \hspace{0.2 cm}
Yes, actually, almost the only places I had been before were the vicinity of
the village where I was born, the concentration camps and the wandering afterwards
in a war-ravaged Europe. Back then, I guess I wasn't really culturally a European. So
it was something new; it wasn't like returning to something. It was my first time in
Paris. The first night that I was there, I went with a colleague, who was also traveling
to the conference and was in Paris for the first time, to a restaurant opposite Notre
Dame. We both ordered fish in a saffron sauce, but neither of us knew what saffron
was. So I looked it up in my French/English dictionary, which just said ``saffron =
saffron''. There wasn't any real translation! We found that really amusing, and enjoyed
the dish.

I didn't stay long - maybe two or three days in Paris, then took the train from
Paris to the conference organized by Prigogine in Brussels, a trip which at that time
took four hours but now takes only one hour! It was a very interesting conference,
where, as a youngster, I asked a lot of questions. We were asked to write down our
questions, which I did. One of the participants in the conference was Robert Brout.
He was an American who went to Brussels and worked with Fran{c}ois Englert, who
recently got the Nobel Prize for joint work with Brout predicting the Higgs boson.
Brout took us out to a restaurant where we ate mussels, which I had never had before,
and then, somehow, I left my notes there, so most of my questions didn't get into
the published proceedings. I remember a picture from that conference; we all looked much
younger at that time! Van Hove was there, and also Giorgio Careri.

\vskip 0.3 cm

L. B. \hspace{0.2 cm}
Actually Careri was already conducting researches that would lead to the demonstration of the existence of quantum vortices in superfluid helium hypothesized by Feynman and Onsager\dots
\vskip 0.3 cm

J. L. \hspace{0.2 cm}
At that conference, I met for the first time Markus Fierz, from Switzerland.
Fierz was a colleague of Wolfgang Pauli and was famous. He was also a delightful
person and an amusing storyteller. I remember him telling a story about Pauli's
father, who was a chemist. His name was Pascheler but, as a young man, he realized
that he had made a mistake in a paper he published. He promptly changed his name to
Pauli and wrote another paper which began with ``Pascheles irrt'' (made a mistake).

\vskip 0.3 cm

L. B. \hspace{0.2 cm}
It's a very nice story, I did not know it! So, after your thesis work, you
applied for a postdoctoral fellowship with Onsager\dots
\vskip 0.3 cm

J. L. \hspace{0.2 cm}
Before getting my PhD, I applied for a fellowship to the National Science
Foundation. In applying for the fellowship you had to say where you wanted to go
and study, and to have somebody specified as a mentor. I wanted to work with Lars
Onsager (who later, in 1968, received the Nobel Prize), and he agreed to be my
mentor.

\vskip 0.3 cm

L. B. \hspace{0.2 cm}
How did your relationship with Onsager begin?
\vskip 0.3 cm

J. L. \hspace{0.2 cm}
I first saw Onsager in my first year as a graduate student at Syracuse. There
was a conference in Pittsburgh on Non-equilibrium phenomena, organized by John
Richardson, and four of us, all students of Peter Bergmann, drove there in my very
old car. I had heard about Onsager relations but I didn't really know what they were.
During the first day of the conference there was a question to the speaker and he said
that maybe professor Onsager could answer that. Then this bald gentleman sitting
right in front of me turned out to be Lars Onsager! So that was how I met him. And
then, a year later, Peter Bergmann organized a conference in Syracuse where I had
a chance to talk with Onsager. I then started reading his papers with S. Machlup on
non-equilibrium statistical mechanics. This was different from the Onsager relations
for which he received a Nobel prize, but also involved the study of fluctuations. I am
referring to the Onsager-Machlup theory \cite{OnsagerMachlup1953}.

When I applied for a post-doc with Onsager in 1956, I was hoping to work with
him on these non-equilibrium problems. Unfortunately he wasn't interested in that
topic at that time. He was working in collaboration with Oliver Penrose on superfluid
helium, which I didn't know anything about. So I had to catch up on that and it was
interesting, but didn't lead to much. We did one paper on this together, which was
published in a conference proceedings \cite{LebowitzOnsager1958}.

In compensation I got to know Onsager as a person not just as a great scientist.
As a post-doc with Onsager I lived in New York and would go to New Haven usually
one day a week and spend the whole day with him. The trip took about two hours.
I would usually arrive at Onsager's office in Sterling laboratory, a long narrow office
with a very tall ceiling and one window at the end, around 11 AM. We would start
talking about some topic, then go out for lunch, frequently to Lars' home. This usually
included some martinis, made expertly by him. After lunch we would meet again in
his office and talk until about 5 PM. (At that time I would usually return to New
York or occasionally stay overnight in New Haven with Roger Vlamsley, a Physics
graduate student who was doing low temperature experiments using liquid helium.)
So while I wasn't doing at Yale what I intended to do and so I didn't benefit as much from that year as I had hoped I learned quite a lot from listening to Onsager talking about all kinds of problems that he was interested in or that people wrote to him about. Getting to know the low temperature experimentalists at Yale, then one of the centers of such work, was another benefit of my fellowship there.

Most important of all was my introduction to Oliver Penrose. He was a post doc
of Onsager during the 1955-1956 academic year. When I first went to Yale in the fall
of 1955 to ask Onsager to be my mentor on the NSF fellowship, he drove me to the
place where Oliver and Joan Penrose lived. There I had my first sherry and started a
lifelong friendship with the Penroses. My scientific collaboration with Oliver, which
has resulted in many joint publications, was one of the most important for me.

\section{Hard spheres}
During the spring of 1957, I started looking for a permanent position. For family
reasons I restricted my search to the New York area. I ended the search by accepting an
assistant professorship in Physics at Stevens Institute of Technology in Hoboken, New
Jersey, just across the Hudson river from mid-town Manhattan. I was able to drive
there in about 20-30 minutes from my apartment in the north of Manhattan. This was
my first teaching experience: teaching Physics to undergraduate engineering students.
I cannot say that I liked teaching, but the Physics Department at Stevens was a nice and lively place. It had on its faculty two former students of Peter Bergman, James Anderson and Ralph Schiller, who were working on relativity, as well as George Yevick, a collaborator of Jerry Percus who had just moved from Stevens to New York University, with whom I collaborated a lot later.

At Stevens I did not work on helium, but started working on my own in both equilibrium
and non-equilibrium statistical mechanics. One of my first works at Stevens was on what later became known as the scaled particle theory of fluids. I did this together
with Harry Frisch and Howard Reiss \cite{ReissFrischLebowitz1959}. I first
met Harry Frisch when I was a graduate student at Syracuse. He spent a year (or
two) there as a postdoc of Peter Bergmann. Harry then moved to Bell Labs, where
he worked with Howard Reiss. Bell Laboratories was at that time one of the foremost
research institutes in the world.

Together with Howard and Harry we developed the scaled particle theory. This is
an approximate theory of the equation of state for a fluid of hard spheres. It turned
out to give a very accurate equation of state compared with machine-computations.
In fact our equation of state was accurate in three dimension up to a density $\rho \sim \frac{1}{3}\pi D^3$, where $D$ is the diameter of the hard spheres. The volume occupied by the balls in then half of the total volume of the container. Around that density, machine
computations show that hard spheres undergo a transition from a fluid phase to a
crystalline phase. Our scaled particle theory did not predict this phase transition, but
as noted before, it gave a very accurate equation of state all the way close to where the
transition takes place. So that was quite interesting and our work was much cited.
Machine computations using molecular dynamics and Monte Carlo methods were
still in an early stage at that time. There was in 1957 a conference in Seattle where
a distinguished panel of theorists took a vote on whether to believe in the existence
of this phase transition. The vote was evenly split\dots

\vskip 0.3 cm

L. B. \hspace{0.2 cm}
Later there was an extension to real fluids of the technique you had applied to rigid sphere systems \cite{HelfandReissFrischLebowitz1960}\dots
\vskip 0.3 cm

J. L. \hspace{0.2 cm}
Yes. There is a whole series of papers by us and by many others applying
these ideas to real fluids. Later it turned out that a totally different approach gave an
identical equation of state for hard spheres. This is something called the Percus-Yevick
equation (PYE) for classical fluids. It is an integral equation for the radial distribution
function $g(r)$ of hard spheres which was solved exactly in the early 1960s by Michael
Wertheim and also by E. Thiele. There are two ways of getting the equation for the
pressure of a fluid once you know $g(r)$. One way is to use the virial theorem and
the other one is to use the compressibility relation. Now, an exact $g(r)$ would give
the same answer for both ways but an approximate $g(r)$, like the one obtained from
the PYE, does not give the same answer. The equation of state coming from the
compressibility gave the same result as scaled particle theory. I still do not know why.

\begin{figure}[ht]
\captionsetup{width=1\textwidth}
\centering
\includegraphics[scale=1.2]{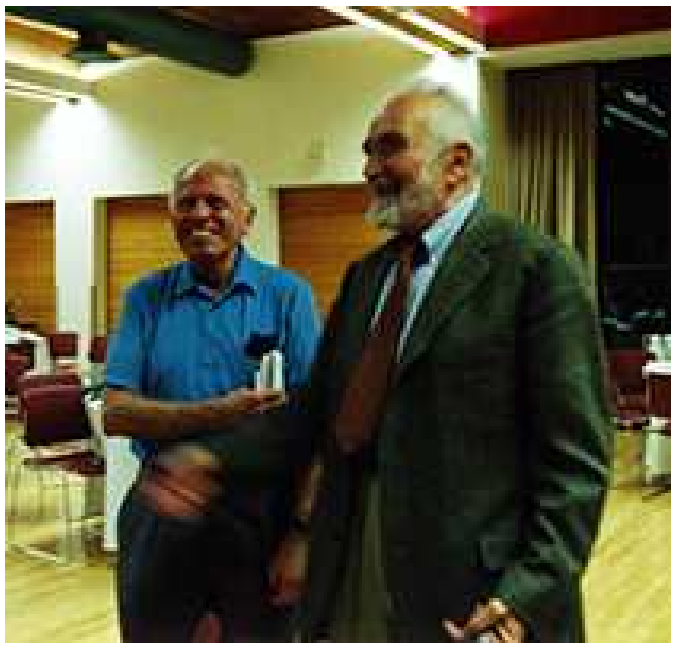}
%\resizebox{0.75\columnwidth}{!}{%
%  \includegraphics{fig1.eps} }
\caption{Joel Lebowitz and Jerome Percus.}
\label{fig:Percus}       % Give a unique label
\end{figure}

Later, in 1959 when I was no longer at Stevens, but had moved to Yeshiva University
in New York, I was able to solve exactly the PYE for a mixture of hard spheres
of different diameters \cite{Lebowitz1964}.

I first found that solution while I was at the annual Gordon Research Conference
on Liquids. It was the first time I had been to such a conference and I was supposed
to speak at the Tuesday evening session. Late Tuesday afternoon I was still working
on that equation. In fact my wife had to go out and get sandwiches both for me
and my postdoc Eigil Praestgaard as we could not stop working to eat dinner in the
dinning hall. I was struggling with this horrible thing, the ratio of something like
a polynomial with 791 terms in the numerator and 794 terms in the denominator.
Suddenly it simplified, a simple denominator with a cubic polynomial for the equation
of state. So I was able to present it that same evening. That was very exciting!

\vskip 0.3 cm

L. B. \hspace{0.2 cm}
I can imagine that your exact solution must have had quite an impact at the time!
\vskip 0.3 cm

J. L. \hspace{0.2 cm}
Yes, this exact solution had a very big impact in that particular community.
In fact another speaker at the same meeting was John Rowlinson, a well-known
chemist from Oxford University, and he had only been able to obtain the first-order
term in the density expansion of the pressure, while I got the full solution.

\vskip 0.3 cm

L. B. \hspace{0.2 cm}
That same year you published with J.S. Rowlinson a paper where you
used the exact solution you had recently obtained \cite{LebowitzRowlinson1964}.
\vskip 0.3 cm

J. L. \hspace{0.2 cm}
In this paper we worked out the thermodynamic predictions of the P.Y.
Equation for mixtures of hard spheres of different sizes. As in the case of the one
component system there was a discrepancy between the results obtained from the
virial and the compressibility relations. The results obtained from the compressibility relation again agreed with that obtained from scaled particle theory. While both agreed with each other and with
machine computations at low and moderate densities (at least to within a few percent)
neither gave any hint of a phase transition predicted by machine-computations. The
mathematical proof of such hard sphere phase transitions is still very much an open
problem about which I continue to think a lot, without any success so far. In any
case the works on scaled particle theory and the PYE brought me in contact with
problems in the theory of classical fluids which I continued to work on for many years
with Percus and other collaborators after I moved to Yeshiva University (where I
stayed for 18 years). Even now, fifty years later, I am still working and hoping for
more mathematical insight into the hard sphere phase transition.

\section{Time Symmetry and Collapse of the Wave Function}

L. B. \hspace{0.2 cm}
In the fall of 1959, you moved from Stevens to Yeshiva University, in New York\dots
\vskip 0.3 cm

J. L. \hspace{0.2 cm}
Yeshiva University (YU) was founded primarily as a school for teaching
both advanced Jewish and secular subjects to orthodox Jewish boys. It therefore had
both a rabbinical and a secular college: students studied the Torah and Talmud in
the mornings and took regular college courses in the afternoon. After some time the
programs were expanded; among other additions, a graduate school of science was
started in 1958-59. The founding dean of that school was Abe Gelbart, formerly a
professor of mathematics at Syracuse University. There I had taken a course with
him and also knew him as a friend of my thesis advisor, Peter Bergmann. Gelbart
invited me to move to Yeshiva as the first faculty member of the new graduate school
in Physics. I accepted the invitation and helped build up the Physics section of what
later came to be named the Belfer Graduate School of Science of Yeshiva University.
I stayed at Belfer until 1977, when it was more or less closed down for lack of funds.
While it lasted, Belfer was an exciting place, with many distinguished visitors. These included Paul Dirac, who spent a year there, during which I got to know him and his wife Margit quite well.  Freeman Dyson also spent a semester there. Regular faculty members included Yakir Aharonov, Peter Bergmann, David Finkelstein, Elliott Lieb, Daniel Mattis\dots  It is well described in the book {\it The Cosmic Landscape} by Lenny Susskind, who was a colleague there for a few years \cite{Susskind2005}.

\begin{figure}[ht]
\captionsetup{width=0.8\textwidth}
\centering
\includegraphics[scale=0.6]{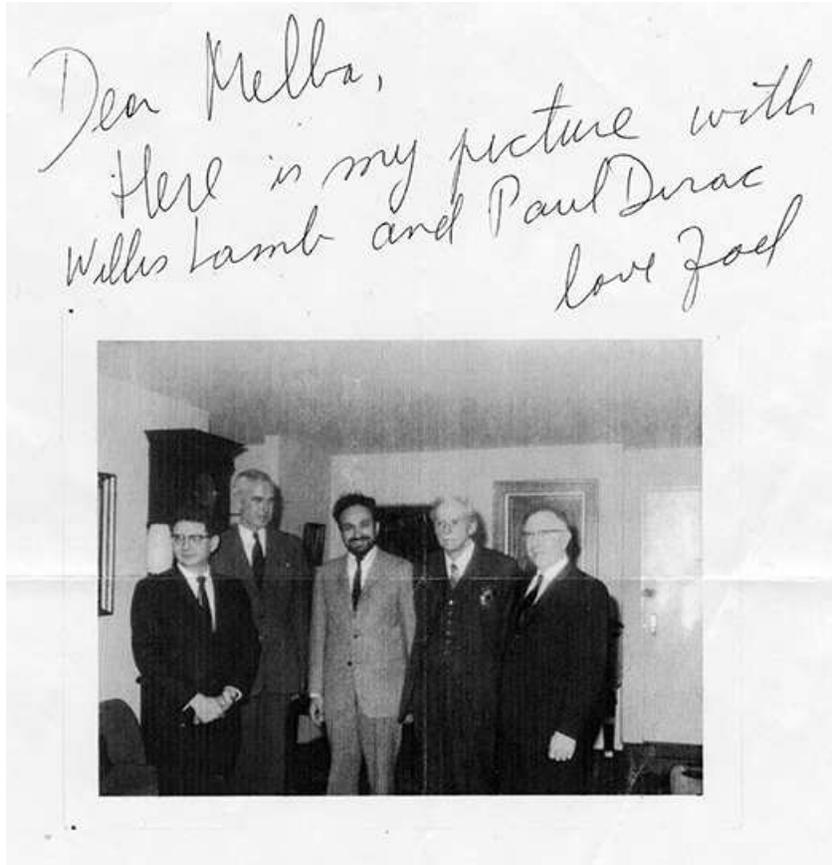}
%\resizebox{0.75\columnwidth}{!}{%
%  \includegraphics{fig1.eps} }
\caption{Dan Mattis, Willis Lamb, Joel Lebowitz, Paul Dirac, Abe Gellart at Yeshiva University. The writing on top addresses Melba Phillips, who was a good friend of Lamb.}
\label{fig:group}       % Give a unique label
\end{figure}

One of the first new works I did at YU was a paper with Yakir Aharonov and Peter
Bergmann on time symmetry in the quantum theory of measurement \cite{AharonovBergmannLebowitz1964}. According to the Copenhagen interpretation of quantum mechanics a system goes into
one of the eigenstates of the quantity measured. Thus if you measure the energy then
it goes into an energy eigenstate corresponding to the value obtained in the measurement.
This so-called collapse of the wave function appears at first as something
very time asymmetric. But we argue that there isn't truly a time asymmetry inherent
in the theory: In principle you could formulate quantum mechanics in a time symmetric
way. The actual asymmetry observed in nature, expressed in the second law
of thermodynamics, comes from the time asymmetric behavior of macroscopic objects
as formulated by Maxwell, Kelvin and Boltzmann. Measurement involves such
macroscopic objects, and this leads to the apparent time asymmetrical collapse.

This led to the idea that between two measurements you could think of the wave
function of the system as being either the one coming from the previous measurement,
which is the usual interpretation of the time asymmetry, or as coming from the later measurement going in the opposite direction. So you can have a kind of time symmetrical superposition of the two wave functions.

Aharonov has been developing and exploiting this idea in various places. He and
some colleagues had a few years ago a {\it Physics Today} article in which they described
their work developing this idea \cite{AharonovPopescueTollaksen2010}. In fact that
joint paper with Aharonov and Bergmann was quite influential\dots
\vskip 0.3 cm

L. B. \hspace{0.2 cm}
So these are still much debated problems at a fundamental level in quantum
mechanics. Actually Wigner already in 1931 had tackled the problem of time
reversal in his fundamental book on the application of group theory to quantum
mechanics. But what was the origin of your work with Bergmann and Aharonov?
\vskip 0.3 cm

J. L. \hspace{0.2 cm}
At the time we were all at the Belfer Graduate School of Science, which
consisted then of one  floor of an old building so we had plenty of opportunities for
discussions. Yes this work owes very much to us being at the same place at the same
time. There is a very nice description of the ambience at Yeshiva University at that
time in the book by  Susskind \cite{Susskind2005}.
\vskip 0.3 cm

\begin{figure}[ht]
\captionsetup{width=0.8\textwidth}
\centering
\includegraphics[scale=0.6]{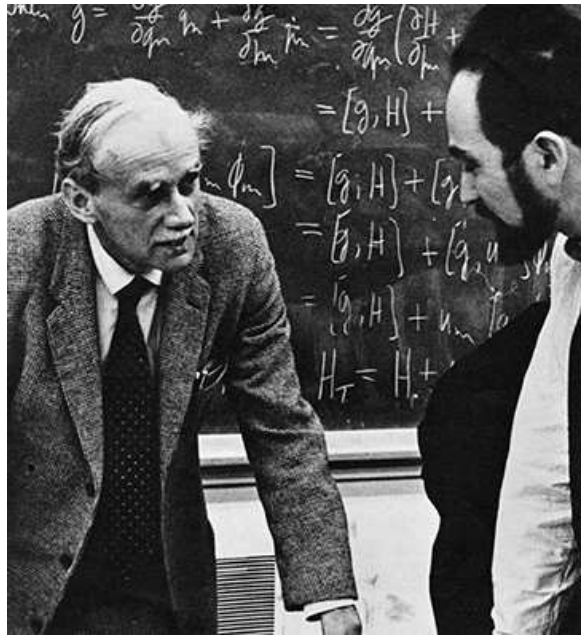}
%\resizebox{0.75\columnwidth}{!}{%
%  \includegraphics{fig1.eps} }
\caption{Joel L. Lebowitz with Paul Dirac.}
\label{fig:Dirac}       % Give a unique label
\end{figure}

\section{The Vapor-Liquid Phase Transitions}

L. B. \hspace{0.2 cm}
And then, in 1964, you began to work with Oliver Penrose\dots
\vskip 0.3 cm

J. L. \hspace{0.2 cm}
As I already mentioned, I had met Oliver Penrose in 1955, when I went
to see Onsager about being a post-doc with him. Then we got to be friends and
after I moved to Yeshiva University Oliver came to spend a year there. At that time
there was a lot of activity, in particular by Kac, Uhlenbeck and Hemmer, (KUH), on
how to derive the well known van der Waals equation of state for a liquid from first
principles \cite{KacUhlenbeckHemmer1963a,KacUhlenbeckHemmer1963b,KacUhlenbeckHemmer1964}. They started by considering
a one-dimensional system which had, in addition to a hard core interaction also a
very long-range attractive potential. That potentials range, was characterized by a
parameter $\gamma$ which was both the inverse length and strength of that potential. They
considered a specific pair potential of the form $\gamma e^{-\gamma r}$ where $r$ is a distance
between a pair of particles. They were able, following some earlier work by Kac, to
compute explicitly the equation of state of that system. For any finite $\gamma$
 there is no phase transition, as there should not be in one dimension, for any sufficiently rapidly
decaying pair potential. However, when they took the limit of $\gamma$ going to zero --- which means they made the potential infinite range they did find a gas-liquid phase
transition. In fact they recovered the old van der Waals equation of state, corrected
by the Maxwell area construction. This was a real mathematical breakthrough. 

\begin{figure}[ht]
\captionsetup{width=0.8\textwidth}
\centering
\includegraphics[scale=0.6]{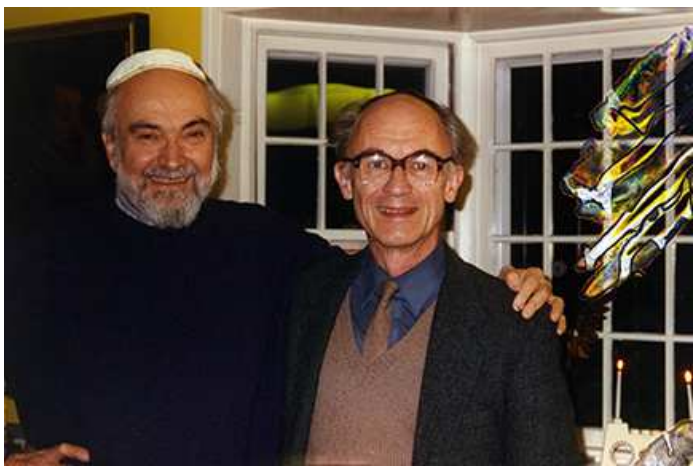}
%\resizebox{0.75\columnwidth}{!}{%
%  \includegraphics{fig1.eps} }
\caption{Joel Lebowitz and Oliver Penrose.}
\label{fig:group}       % Give a unique label
\end{figure}

What
Penrose and I did was to generalize that work by KUH to arbitrary dimensions and
a large class of long-range and short-range potentials. We proved that in the limit $\gamma \to 0$ this long-range potential will give a liquid vapor transition of the van der
Waals type with Maxwell's equal area construction \cite{LebowitzPenrose1966}.

This work has been greatly extended by various authors. In particular there is a
whole book on this subject by Errico Presutti from Rome, {\it Scaling Limits in Statistical
Mechanics and Microstructures in Continuous Mechanics} \cite{Presutti2009}.

\begin{figure}[ht]
\captionsetup{width=0.8\textwidth}
\centering
\includegraphics[scale=0.95]{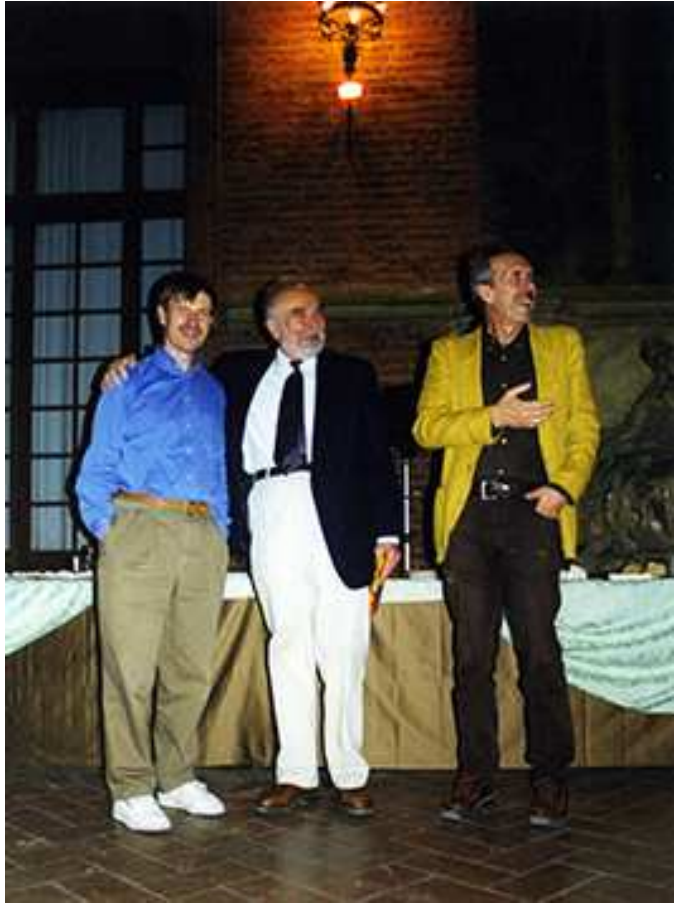}
%\resizebox{0.75\columnwidth}{!}{%
%  \includegraphics{fig1.eps} }
\caption{Herbert Spohn (left), Joel Lebowitz and Errico Presutti.}
\label{fig:Dirac}       % Give a unique label
\end{figure}

This book also describes related work done some years later, by Presutti, Alex
Mazel and myself \cite{LebowitzMazelPresutti1999}. In that work we were able to extend
the work with Penrose and prove, in some more specialized cases, and dimension
2 or more, that if $\gamma$ is small enough, i.e. when the range of the attractive potential is
large enough, then the system will undergo a phase transition, i.e. it is not necessary
that $\gamma$ go to zero. This extended to continuum systems what had been previously
proven only for lattices systems; namely the existence of phase transitions in systems
with a finite range interaction without any symmetry.

The first rigorous proof of a phase transition was given by Onsager for the Ising
model which does have a symmetry \cite{Onsager1944}. Following Onsager's breakthrough Pirogov and Sinai proved the existence
of phase transitions for a large class of lattice systems without symmetry at low temperature \cite{PirogovSinai1975,PirogovSinai1976}.
This is not the case when you have continuum systems. The cases where one can
prove phase transitions is very, very limited. There is the quantum mechanical Bose-Einstein condensation but this is only proven for ideal gases.

For classical systems there is a result by Ruelle of a phase transition in the Widom-Rowlinson model of two species of particles, call them A and B  \cite{Ruelle1971}. In this model
particles do not interact with particles of their own species, but there is a hard core
repulsion between particles of species A and species B. When you increase the density
of the gas having both A's and B's the system will segregate into a A-rich and a
B-rich phase leading to a phase transition similar in many ways to what happens in
the Ising model.

Ruelle's proof uses heavily the symmetry of the problem. We still do not have a
proof of there being a phase transition for ordinary fluid systems with rapidly decaying
interactions, such as those described by Lennard-Jones potentials. Such systems show
phase transitions in computer simulations and the simulations agree with what is
observed in experiments on real fluids, like Argon or Neon, where quantum mechanics
does not play an important role. So the work with Mazel and Presutti is interesting\dots

Following our rigorous derivation of the liquid-vapor phase transition for systems with long range Kac potentials, Penrose and I gave a precise definition of metastable states for such systems. These involved defining a region in the phase space of the system where the density is constrained to be approximately uniform with the same average density in boxes of size $ 1/\gamma$. This prevents the system from segregating into a liquid and a vapor phase. This is possible because the range of the Kac potential, causing the liquid-vapor transition is very large compared to the distances between the particles and any short-range interaction, but still very small compared to the size of the system \cite{PenroseLebowitz1971}. 

We also showed that when the Kac potential is positive (rather than attractive), then a phase transition caused by any short range potential with segregated phases will turn into a ``foam'' with a mixture of the two phases. 

%\section{Collaboration with Elliot Lieb}
\vskip 0.3 cm

L. B. \hspace{0.2 cm}
Your work with Penrose \cite{LebowitzPenrose1966} was immediately cited by Elliott Lieb \cite{Lieb1966} in his article extending your proof to quantum systems with any statistics --- Boltzmann, Bose or Fermi-Dirac\dots And the first joint paper you and Lieb wrote together with Burke \cite{LebowitzBurkeLieb1966}, looks like an immediate follow-up of Liebs's and Lebowitz-Penrose's articles\dots
\vskip 0.3 cm

J. L. \hspace{0.2 cm}
Lieb was for a short time my colleague at Yeshiva University and so we did some work together there and also after that. This will be described in Part 2 of this article\dots

\begin{figure}[ht]
\captionsetup{width=0.8\textwidth}
\centering
\includegraphics[scale=0.6]{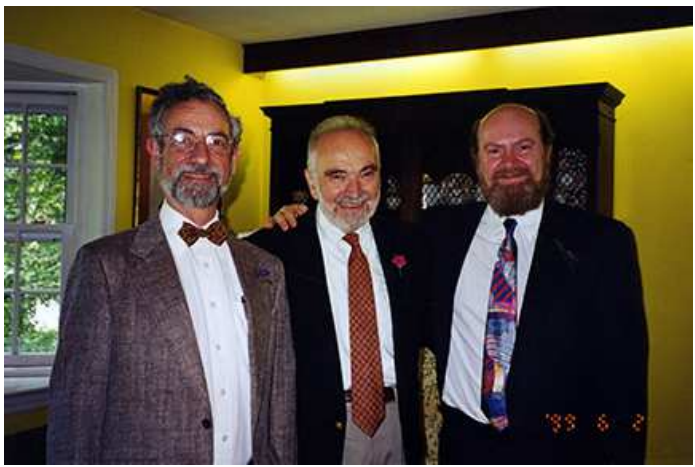}
%\resizebox{0.75\columnwidth}{!}{%
%  \includegraphics{fig1.eps} }
\caption{Joel Lebowitz with Elliott Lieb (left) and Michael Aizenman.}
\label{fig:Lieb}       % Give a unique label
\end{figure}

\section{Acknowledgments}
We are deeply indebted to Wolf Beiglb\"ock for a critical and constructive evaluation of the manuscript and for his and Francesco Guerra's constant encouragement. J. L. L. thanks the US AFOSR and NSF for their long time support of his research. We also gratefully acknowledge Sabine Lehr's careful transcription of the interview.

\end{document}